# Enhancement of Kv1.3 Potassium Conductance by Extremely Low Frequency Electromagnetic Field

C. Cecchetto[1,2], M. Maschietto[1], P. Boccaccio[3], S. Vassanelli[1,*]

[1]*University of Padova, Italy, Dept. of Biomedical Sciences, via Marzolo 3, Padova, Italy,*

[2]*University of Padova, Italy, Dept. of Information Engineering, via Gradenigo 6/B, Padova, Italy*

[3]*Istituto Nazionale di Fisica Nucleare, Laboratori Nazionali di Legnaro, Legnaro, Padova, Italy*

*Correspondence to: Prof. Stefano Vassanelli, Dept. of Biomedical Sciences, NeuroChip Lab, Via Francesco Marzolo 3, 35131, Padova, Italy. E-mail: stefano.vassanelli@unipd.it

## Abstract

Theoretical and experimental evidences support the hypothesis that extremely low-frequency electromagnetic fields can affect voltage-gated channels. Little is known, however, about their effect on potassium channels. $K_v1.3$, a member of the voltage-gated potassium channels family originally discovered in the brain, is a key player in important biological processes including antigen-dependent activation of T-cells during the immune response. We report that $K_v1.3$ expressed in CHO-K1 cells can be modulated in cell subpopulations by extremely low frequency and relatively low intensity electromagnetic fields. In particular, we observed that field exposure can cause an enhancement of $K_v1.3$ potassium conductance and that the effect lasts for several minutes after field removal. The results contribute to put immune and nervous system responses to extremely low-frequency electromagnetic fields into a new perspective, with $K_v1.3$ playing a pivotal molecular role.

# INTRODUCTION

Electromagnetic fields in the 0.03 – 300 Hz frequency range are commonly known as extremely low-frequency electromagnetic fields (ELF-EMF) and are found in the public or domestic environment with intensities up to several hundreds of µT near appliances [World Health Organization (WHO), 2015]. Several studies provided evidence that exposure to ELF-EMF can cause biological effects, such as a modulation of cell proliferation and differentiation, but molecular actors behind such cellular responses remain to be identified [Funk et al., 2009]. In mammalian cells, voltage-gated ion channels (VGC) are among suitable candidates. In fact, theoretical models [Funk et al., 2009] and experimental studies on calcium [Pall, 2013] and sodium channels [He et al., 2013] support this hypothesis, with fields that may act on the channel directly or by interposition of cellular signalling pathways [Gartzke and Lange, 2002; Funk et al., 2009; He et al., 2013; Pall, 2013]. Yet, effects on VGC of the potassium family, $K_v$, are mostly unexplored, except for the recent work of [Gavoçi et al., 2013] reporting that an AC low-frequency (8 Hz) and low-intensity (100 µT) field was not altering Tetraethylammonium (TEA)-sensitive potassium currents in neuroblastoma cells.

$K_v$ channels are of paramount importance in cell physiology and represent one of the most diverse and ubiquitous families of membrane proteins. The $K_v$ superfamily comprises 12 subfamilies ($K_v1$-$K_v12$), all formed by tetrameric membrane-spanning protein complexes that create a selective pore for potassium ions. Each subunit is formed by six transmembrane segments (S1-S6), with S1-S4 constituting a voltage sensor domain (VSD) and S5-S6 the pore-forming region [Coetzee et al., 1999; Gutman et al., 2003]. $K_v1.3$ is a member of the *Shaker* ($K_v1$) delayed rectifier family and it was identified in the brain and in several regions throughout the body [Swanson et al., 1990; Bielanska et al., 2009; Bielanska et al., 2010]. In particular, although mainly expressed in the olfactory bulb and lymphocytes [Duque et al., 2013], $K_v1.3$ is found in the hippocampus [Veh et al., 1995], epithelia [Grunnet et al., 2003], adipose tissue [Xu et al., 2004], and both skeletal and smooth muscle [Villalonga et al., 2008; Bielanska et al., 2012a; Bielanska et al., 2012b].

Intriguingly, in addition to the plasma membrane, the channel's presence in mitochondria has been recently demonstrated [Szabo and Zoratti, 2014; Szabò et al., 2010].

$K_v1.3$ has attracted much attention for playing diverse biological functions relevant to clinics. Above all, it is involved in antigen-dependent T-cells activation during the immune response together with the $Ca^{2+}$-activated IKCa1 $K^+$ channel, and is therefore considered a potentially elective therapeutic target for immunomodulation [Beeton et al., 2001; George Chandy et al., 2004; Gocke et al., 2012; Hansen, 2014; Kahlfuß et al., 2014; Panyi, 2005; Wang and Xiang, 2013; Wulff et al., 2003]. Furthermore, it is assumed to contribute to class-switching of memory B cells [Wulff et al., 2004]. Along with such immune system-specific functions, $K_v1.3$ has been also proposed to contribute to modulation of sensitivity to insulin [Upadhyay et al., 2013] and to the regulation of crucial cellular processes such as apoptosis [Gulbins et al., 2010; Szabò et al., 2010] and proliferation [Bielanska et al., 2009].

In the present work, we investigated the effect of ELF-EMF on $K_v1.3$ expressed in CHO-K1 cells. We focused on whole-cell conductance, which was found to be enhanced in cell subpopulations by relatively low-intensity fields in the order of hundreds of µT.

## MATERIALS AND METHODS

### Cell culture and $K_v1.3$ channel transfection

CHO-K1 cells (ATCC, USA) were cultured in 35 mm Petri dishes (density: $2 \times 10^3$ cells/cm$^2$) and maintained in F-12 Nutrient Mixture – Ham – supplemented with 10% (v/v) heat-inactivated FBS, 10 u/ml penicillin and 10 µg/ml streptomycin, in a humidified 5% $CO_2$ atmosphere and at constant temperature (37 °C). Cells were co-transfected with EGFP and $K_v1.3$ plasmids (2 µg each), kindly provided by Prof. P. Fromherz (Max-Planck Institute for Biochemistry, Martinsried, Germany) [Kupper, 1998]. One to two days from transfection, fluorescent cells were pre-selected for electrophysiology (see below). Plasmids were amplified and purified from bacterial strains

(HiSpeed Plasmid Maxi Kit, QIAGEN, Italy) and transfected with Lipofectamine™2000 one to two days after plating. If not otherwise indicated, reagents and culture media were purchased from Life Technologies (Italy).

**Electrophysiology with ELF-EMF exposure**

The electromagnetic field was generated by a 20 Hz sinusoidal current (provided by a function generator Agilent 33220A, Agilent Technologies Inc.) applied to a custom made solenoid, formed by an enamelled copper wire (0.45 mm wire diameter, 2000 turns/metre) wound on a plastic support (80 mm diameter, 10 layers of wires). The cell culture in the Petri dish was placed at the centre of the solenoid and exposed to a magnetic field orthogonal to the plane of adhesion (Figure 1). Two field intensities were employed as detailed in the Results section: ELF-EMF$_1$ = 268 µT and ELF-EMF$_2$ = 902 µT (r.m.s. values measured with a Gauss meter Model 907, Magnetic Instrumentations Inc., Indianapolis, IN, USA). While for ELF-EMF$_1$ the function generator was driving directly the solenoid, an amplifier was used to boost the signal to generate ELF-EMF$_2$. The value of 20 Hz was chosen basing on previous observations [Medlam et al., 2008; Medlam, 2009; Cecchetto et al., 2013; Cecchetto et al., 2014]. Experiments were performed at room temperature, with the extracellular recording solution (see below) monitored by a digital thermometer (GTH 1150, Greisinger).

$K_v1.3$ potassium currents were recorded 24-48 h after transfection by patch-clamp in whole-cell voltage-clamp configuration. Recordings were performed before, during and after field exposure using an Axopatch 200B amplifier (Molecular Devices, USA) connected to the PC through a BNC-2110 Shielded Connector Block (National Instruments Corp, Austin, TX, USA) along with a PCI-6259 PCI Card (National Instruments Corp, Austin, TX, USA). WinWCP (Strathclyde Electrophysiology Software, University of Strathclyde, Glasgow, UK) was used for data acquisition and data analysis. Raw traces were low-pass filtered at 3.3 kHz and sampled at 10 kHz. Micropipettes were pulled from borosilicate glass capillaries (GB150T-10, Science Products GmbH, Hofheim, Germany) using a P-97 Flaming/Brown Micropipette Puller (Sutter Instruments Corp., Novato, CA, USA). Extracellular and intracellular solutions contained, respectively (in mM): 135.0 NaCl, 5.4 KCl, 1.0 MgCl2, 1.8 CaCl, 10.0 glucose, 5.0 HEPES (adjusted to pH 7.4 with 1N NaOH); 140.0 KCl, 2.0 MgCl2, 5.0 EGTA, 5.0 HEPES (adjusted to pH 7.3 with 1N KOH). The resistance of the micropipettes in extracellular solution, when filled with intracellular solution, ranged from 2 to 3 MΩ. Current transients due to capacitive (stray, pipette and membrane) components were cancelled through the amplifier compensation circuits, whereas leakage currents were subtracted using a P/4 protocol. The voltage clamp protocol consisted of a depolarizing test pulse of +100 mV amplitude from a –90 mV holding potential for 200 ms, thus obtaining, within the pulse duration, a full $K_v1.3$ activation [Grissmer et al., 1994; Marom and Levitan, 1994]. Depolarizing pulses were applied, with a few exceptions as detailed in the text, at time intervals > 30 s to avoid cumulative inactivation of the channel [Cook and Fadool, 2002; Grissmer et al., 1994; Marom and Levitan, 1994]. The steady-state potassium conductance for activation, $g_K$, was computed from potassium currents (see below) and estimating the potassium reversal potential from intra- and extracellular recording solutions using the Nernst equation. During exposure, a 20 Hz modulation artefact was seen in the electrophysiological trace (see **Figure 2B**, exposure trace). Thus, as the artefact was masking the peak potassium current, $g_K$ was estimated from the potassium current averaged across the last full sinusoidal period (50 ms duration) during the test pulse. Similarly, in the absence of the field, currents were averaged across the last 50 ms of the pulse. As such, the measures represented approximations to the $K_v1.3$ steady-state conductance after full channel activation while neglecting an error introduced by slow inactivation [Grissmer et al., 1994; Marom and Levitan, 1994]. Statistical analysis was performed using a GraphPad Prism (GraphPad Software Inc., Version 5.01) as detailed in the text.

## RESULTS AND DISCUSSION

The effect of ELF-EMF on $K_v1.3$ was investigated by exposing $K_v1.3$ expressing CHO-K1 cells to a field and monitoring changes of the steady-state potassium current by patch-clamp in whole-cell voltage-clamp configuration. Potassium currents were activated throughout the experiment by applying depolarizing test pulses of 200 ms duration to +10 mV and then computing the corresponding steady-state conductance, $g_K$, as described in 'Materials and Methods'. Pulses were repeated to record currents prior, during and after cell exposure to ELF-EMF, the field being applied one minute after entering the whole cell configuration, held for 3 min and then removed. $K_v1.3$ currents recorded from one representative cell exposed to the ELF-$EMF_2$-type field are shown in **Figure 2**. Figure 2A shows one representative cell during the whole-cell patch-clamp procedure. In Figure 2B three superimposed Kv1.3 current traces recorded under exposure of ELF-$EMF_2$ are reported: the first one recorded 1 min before exposure (labelled as PRE), the second one recorded during exposure (i.e., at 3 min from the onset of field application, labelled as EXPOSURE) and the third one at 1 min after the end of exposure (labelled as POST). In those traces, an increase of the $K_v1.3$ steady-state potassium current can be appreciated during the exposure phase. The increase with respect to pre-exposure conditions was nearly 50 percent which can be recognized despite a 20 Hz sinusoidal modulation artefact obscuring the current trace. Interestingly, such enhancement of the $K_v1.3$ current lasted, in part, even beyond field application. This is exemplified in **Figure 2B**, as seen in the trace recorded 1 min after field removal (the POST trace) where the current intensity is still about 10 percent higher than in the pre-exposure phase.

We performed a statistical analysis of the effect of ELF-EMF on $K_v1.3$ focusing on the evolution of the potassium current response over time and on its dependency on the field intensity. Two cell populations were analysed: one exposed to the lower intensity field, ELF-$EMF_1$, and one to ELF-$EMF_2$ (see Materials and Methods). Percentage conductance changes of the steady-state potassium conductance, $\Delta g_K$ (%), are reported over time for four cells (two exposed to ELF-$EMF_1$ and two to

ELF-EMF$_2$) in **Figure 3A**. Each point in the plots represent the value of $\Delta g_K$ derived from one single depolarizing test pulse. Note that the time interval between subsequent pulses was chosen to be larger than 30 s in order to avoid cumulative inactivation of the channel [Cook and Fadool, 2002; Grissmer et al., 1994; Marom and Levitan, 1994] and consequent misjudgements on the effect of the field. Only in one case the inter-pulse interval was of 10 s, that is, at the transition from the end of exposure to the post-exposure period. In fact, an appreciable conductance drop is seen at this first post-exposure checkpoint in three out of four cells (namely cells 1, 3 and 4), which can likely be ascribed to cumulative inactivation. The plots demonstrate the large variability we encountered across cells both in terms of dynamics of the response and of its extent: cells 1 and 2, exposed to the low intensity field, show a gradual increase of $\Delta g_K$ from the start of exposure, reaching a maximum at the end of field application (3 min from the application onset) and then smoothly decreasing across the following 3 min after field removal. Cells 3 and 4, instead, present an abrupt increase right after field application and then a progressive decrease which continues during the post exposure period. In terms of the effect's extent, $\Delta g_K$ varies between 30 to nearly 240 percent across the four cells.

In spite of these representative examples, we observed a significant number of cells that did not clearly respond to the field, and this under both exposures. Thus, when considering averaged responses over the entire cell populations, only a slight tendency to $\Delta g_K$ increase emerged (**Figure 3B**). Interestingly, however, the time course of $\Delta g_K$ was differing between ELF-EMF$_1$ and ELF-EMF$_2$ treated cells. In the first case, cells were responding with a slow $g_K$ increase reaching a steady-state after about 1 min from the exposure onset; in the second case, the conductance increased suddenly upon field application.

A clearer picture of the effect of ELF-EMF$_1$ and ELF-EMF$_2$ on K$_v$1.3 emerged from the analysis of the population distributions with respect to $\Delta g_K$. In particular, we considered the distributions of $\Delta g_K^{max}$, the maximum $\Delta g_K$ observed irrespective of its time of occurrence within the exposure window, for three different cell populations (**Figure 4**): (i) one exposed to ELF-EMF$_1$, (ii) one exposed to ELF-EMF$_2$ and one not exposed to the field (control). The control, which was assumed having a normal distribution with respect to $\Delta g_K^{max}$, was tested for normality and fitted with a Gaussian curve (least square fit, dark line in **Figure 4**). By comparing ELF-EMF$_1$ and ELF-EMF$_2$ distributions with the normally distributed control, we evidenced how cell exposure resulted in a deformation of the distribution due to the higher number of cells showing a $g_K$ increase under field exposure (**Figure 4**). The effect appeared to be more pronounced for ELF-EMF$_2$ as compared to ELF-EMF$_1$, with the fraction of cells falling beyond 3σ (σ=standard deviation, SD) at the right side of the control curve accounting for a 26% and 37%, respectively. The distribution of $\Delta g_K^{max}$ time occurrences observed in these outliers can be summarized as follows: 21% of cells for ELF-EMF$_1$ and 13% for ELF-EMF$_2$ at 10 s from field application; 33% for ELF-EMF$_1$ and 44% for ELF-EMF$_2$ at 1 min; 46% for ELF-EMF$_1$ and 44% for ELF-EMF$_2$ at 3 min. We also compared the medians of the distributions through a Wilcoxon Signed Rank Test (median equal to zero as null hypothesis) and found that, contrary to the control, medians of ELF-EMF$_1$-exposed and ELF-EMF$_2$-exposed populations differed significantly from zero (p = 0.014 and p = 0.0001, respectively).

Taken together these results provided a strong indication that ELF-EMF can enhance K$_v$1.3 conductance. The effect, however, is distributed unevenly across cells, which led to asymmetry in the frequency distributions reported in **Figure 4**. Increasing field intensity further increases the number of responsive cells. It is noteworthy that the maximum K$_v$1.3 conductance enhancement was frequently found at least 1 min from the start of exposure and the effect was lasting beyond field removal. Although it is difficult to argue about molecular mechanisms on this basis, we may infer that the activation of intracellular signalling pathways of K$_v$1.3 modulation could be involved [Cook and Fadool, 2002; Fadool et al., 2000; Fadool and Levitan, 1998; Bowlby et al., 1997], thus delaying the onset of the response and justifying its persistence in the absence of the field.

Whatever the mechanism is, our findings suggest that ELF-EMF may produce cellular effects through K$_v$1.3 modulation, particularly at the level of the immune and nervous systems. Starting from these observations, and before considering implications for therapy and prevention, additional studies will be required to elucidate mechanisms and assess whether the effect is found also in native Kv1.3-expressing cells.

**Acknowledgments**

We want to thank M. Mahmud for the graphical tips and Prof. D. Pascoli and M. Mahmud for critically reading the manuscript.

**Contributions of authors**

SV and PB designed the experiments. MM prepared the cellular cultures and did the plasmid amplification and transfection. CC performed all electrophysiology experiments and data analyses. SV and CC wrote the manuscript.

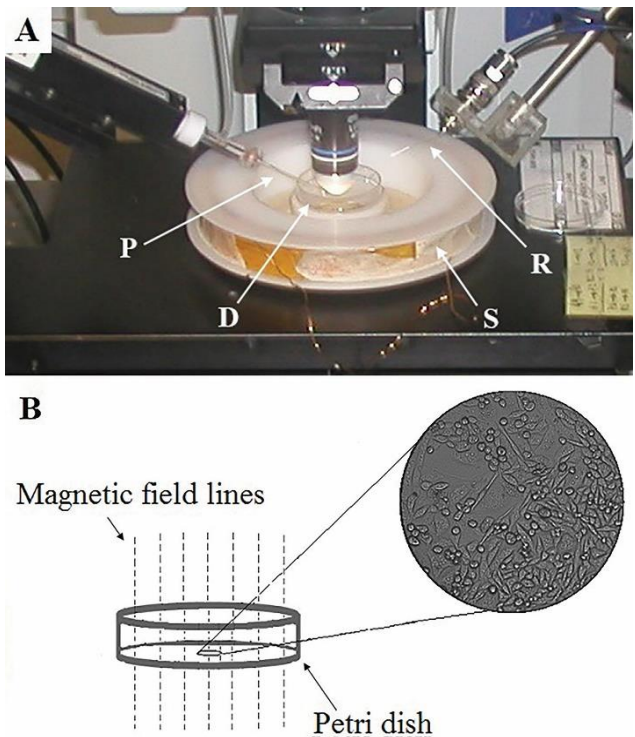

**Figure 1 Experimental setup for ELF-EMF exposure and simultaneous whole-cell patch-clamp. A.** Photograph showing the main components of the setup: the solenoid (S) formed by a copper wire wound on a plastic spool is visible at the foreground with the Petri dish (D) at the centre; a patch-clamp pipette (P) points to a cell beneath the microscope objective; the Ag/AgCl reference electrode (R) is immersed in the recording solution close to the border of the Petri dish. **B**: Sketch of the Petri dish and field direction (dashed lines). Blow up: CHO-K1 cells cultured at the bottom of the dish as observed by bright-field microscopy (Olympus BX51WI microscope, water immersion 10x objective, 0.30 N.A., 6.5 mm W. D., from Olympus Italia SRL, Milan, Italy).

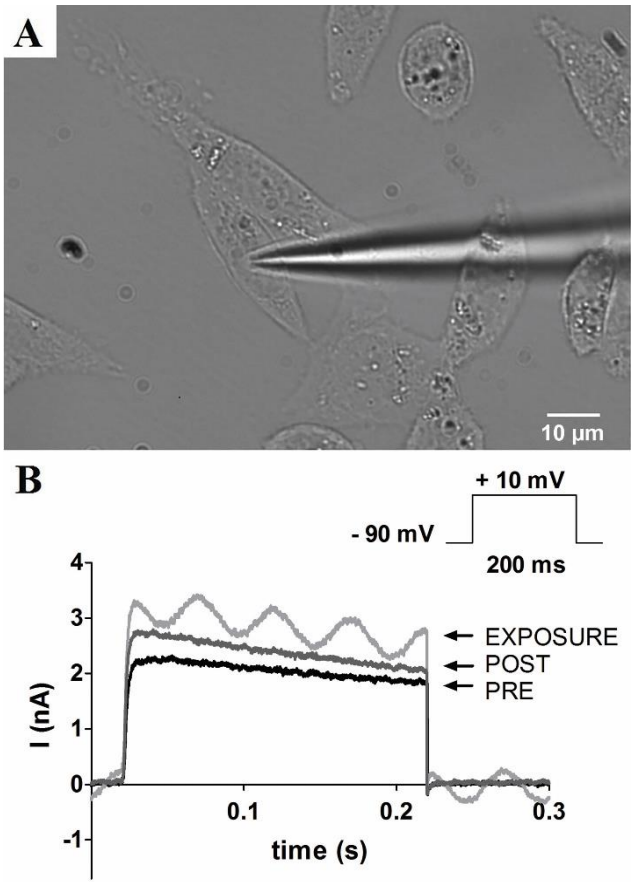

**Figure 2 Effect of ELF-EMF on Kv1.3 potassium current. A**. Bright-field micrograph of a Kv1.3 expressing CHO-K1 cell during an experiment of exposure to the ELF-EMF$_2$ field. The cell was contacted by a patch-clamp pipette for recording potassium currents that are reported in B. **B**: Example of three whole-cell current traces recorded, respectively, at 1 min before exposure (PRE), under field exposure at 3 min from the start of field application (EXPOSURE) and 1 min after the field was removed (POST). A potassium current increase of about 50 percent can be clearly observed in the EXPOSURE trace, together with a 20 Hz AC modulation artefact. The increase persists till 1 min post-exposure, although to a reduced extent with respect to the exposure phase. *Top-right*: voltage-clamp protocol for Kv1.3 current activation.

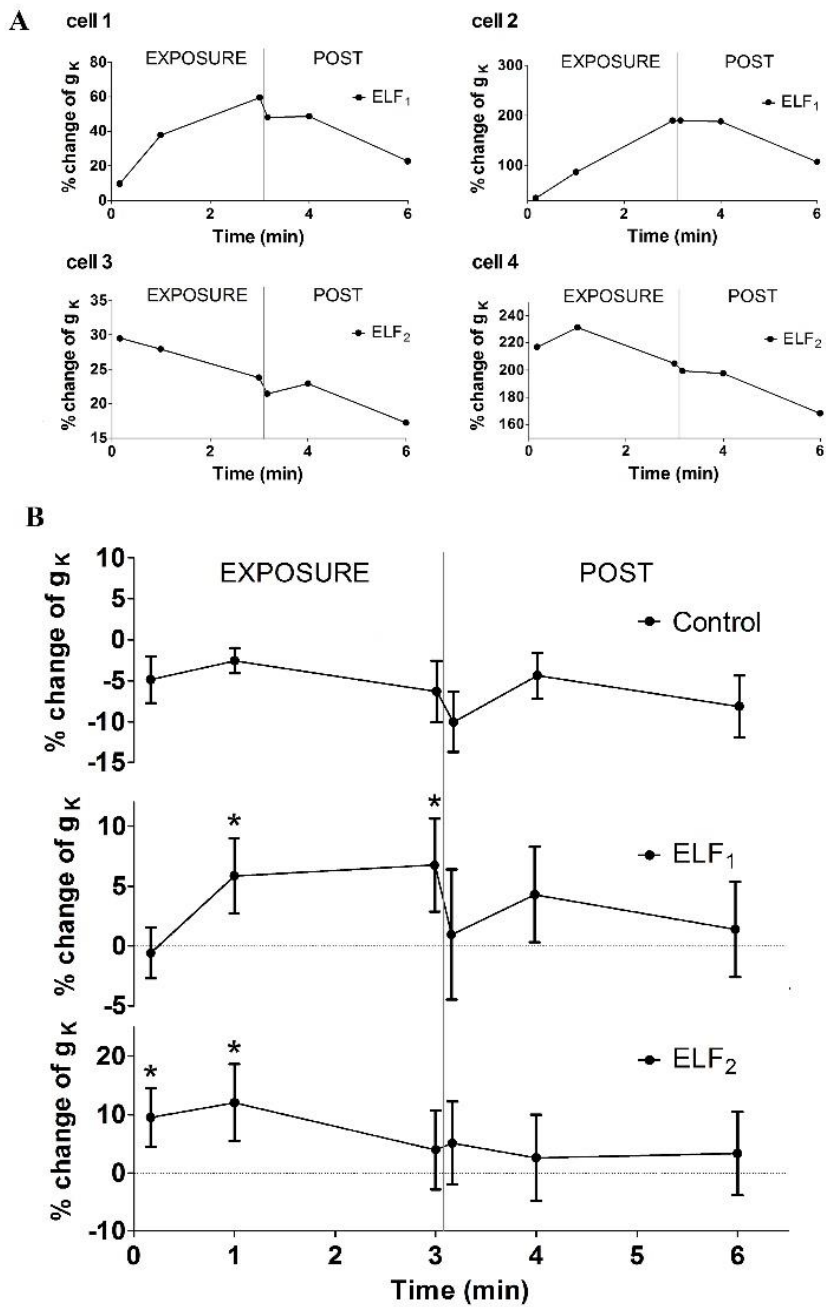

**Figure 3 Average response of $K_v1.3$ steady-state conductance to ELF-EMF$_1$ and ELF-EMF$_2$.** *A:* Response of representative cells (cells 1 and 2 exposed to ELF-EMF$_1$, cells 3 and 4 exposed to ELF-EMF$_2$) in terms of $g_K$ percentage changes with respect to pre-exposure ($\Delta g_K$). *B:* Average $\Delta g_K$ across the population of cells exposed either to ELF-EMF$_1$ (N = 92) or to ELF-EMF$_2$ (N = 44). Data are plotted as mean ± SEM. Only slightly significant increases of conductance were observed in the two populations: at 1 and 3 min of exposure for ELF-EMF$_1$ (p = 0.065 and p = 0.086, respectively) and at 10 s and 1 min for ELF-EMF$_2$ (p = 0.067 and p = 0.074, respectively). Note the different dynamics of the response to the field in the two cell populations with ELF-EMF$_1$-exposed cells showing a gradual rise in conductance and then a plateau, and ELF-EMF$_2$-exposed cells an abrupt conductance increase upon field application followed by a gradual decrease after 1 min of exposure.

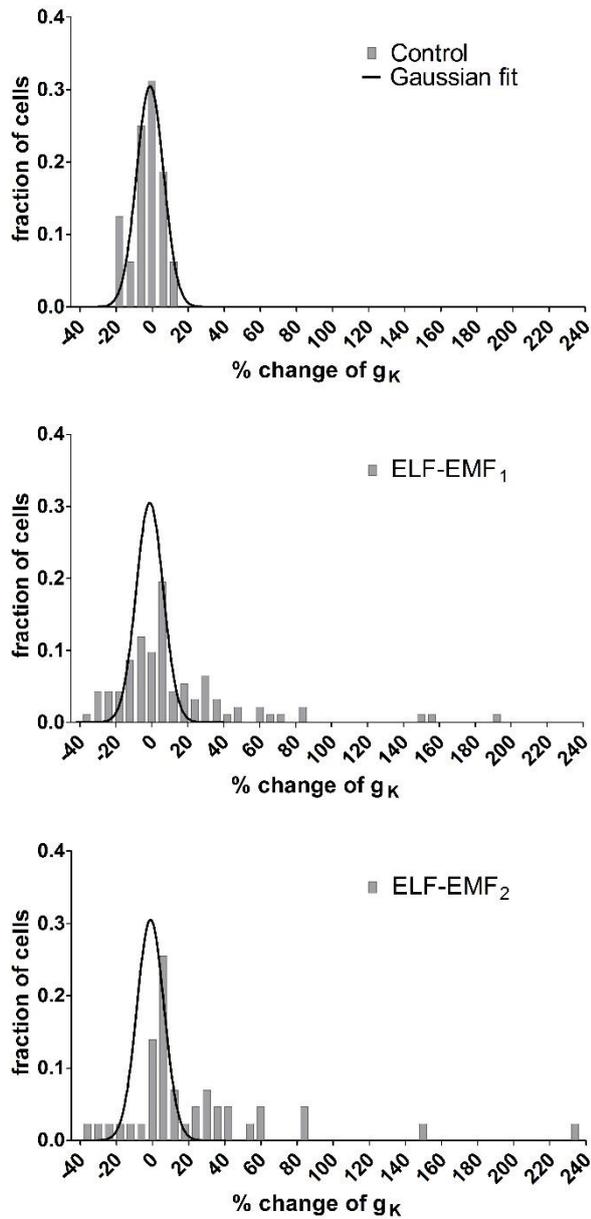

**Figure 4**. **Distributions of maximum percentage $g_K$ changes**. Frequency distributions of $\Delta g_K^{max}$ in control, ELF-EMF$_1$-exposed and ELF-EMF$_2$-exposed cell populations built with a fixed bin equal to 6% (Control: N = 16; ELF-EMF$_1$: N = 92; ELF-EMF$_2$: N = 44. Total number of cells: N = 156). Raw control data were tested for normality using a D'Agostino & Pearson omnibus K2 test (p = 0.89; null hypothesis: non-Gaussian distribution) and fitted with a Gaussian curve (least squares fit, $R^2$=0.937). The same normality test performed on raw data from ELF-EMF$_1$ and ELF-

EMF$_2$ cell populations evidenced, in both cases, a significant deviation from normality (p < 0.0001 for both).